\documentclass{article}
\usepackage{spconf,amsmath,graphicx}
\usepackage{multirow}
\usepackage{float}
\usepackage[fontsize = 9.0pt]{scrextend}
\usepackage{hyperref}
\usepackage{booktabs}
\usepackage{caption}

\title{Speech recognition for air traffic control via feature learning and end-to-end training}
\name{Peng Fan$^{1}$, Dongyue Guo$^{1}$, Yi Lin$^{1,2}$ , Bo Yang$^{1,2}$, Jianwei Zhang$^{1,2}$ \thanks{Yi Lin is the corresponding author. This work was supported by the National Natural Science Foundation of China (No.62001315). }}
\address{$^{1}$National Key Laboratory of Fundamental Science
on Synthetic Vision, Sichuan University, Chengdu 610065, China, \\
$^{2}$College of Computer Science, Sichuan University, Chengdu 610065, China}

\begin{document}
%
\maketitle
\begin{abstract}
In this work, we propose a new automatic speech recognition (ASR) system based on feature learning and an end-to-end training procedure for air traffic control (ATC) systems. The proposed model integrates the feature learning block, recurrent neural network (RNN), and connectionist temporal classification loss to build an end-to-end ASR model. Facing the complex environments of ATC speech, instead of the handcrafted features, a learning block is designed to extract informative features from raw waveforms for acoustic modeling. Both the SincNet and 1D convolution blocks are applied to process the raw waveforms, whose outputs are concatenated to the RNN layers for the temporal modeling. Thanks to the ability to learn representations from raw waveforms, the proposed model can be optimized in a complete end-to-end manner, i.e., from waveform to text. Finally, the multilingual issue in the ATC domain is also considered to achieve the ASR task by constructing a combined vocabulary of Chinese characters and English letters. The proposed approach is validated on a multilingual real-world corpus (ATCSpeech), and the experimental results demonstrate that the proposed approach outperforms other baselines, achieving a 6.9\% character error rate.
\end{abstract}
\section{Introduction}
\label{sec:intro}

Automatic speech recognition (ASR) can translate speech into computer-readable texts \cite{el2011survey}. In air traffic control (ATC), radio speech is the primary way of communication between air traffic controllers (ATCo) and pilots. The ASR is introduced into the ATC system to translate the speech of the ATCo and the pilot, which can be used to reduce the workload on the ATCo and ensure flight safety \cite{geacuar2010reducing}.

Compared to the common ASR research, the ATC has many new challenges and difficulties. In general, the ATCo and pilots speeches are usually in English. However, in China, the ATCos and pilots communicate through Chinese for the domestic flight more frequently. That is to say, speech on the same frequency usually in both Chinese and English, i.e., multilingual ASR is required for the ATC domain \cite{lin2020unified}. Our previous work introduced ASR into the ATC safety monitoring framework, and also converted ATCo and pilot speech into instructions for controlling intent inference \cite{lin2019real}.

\begin{figure*}[h!]
\centerline{\includegraphics{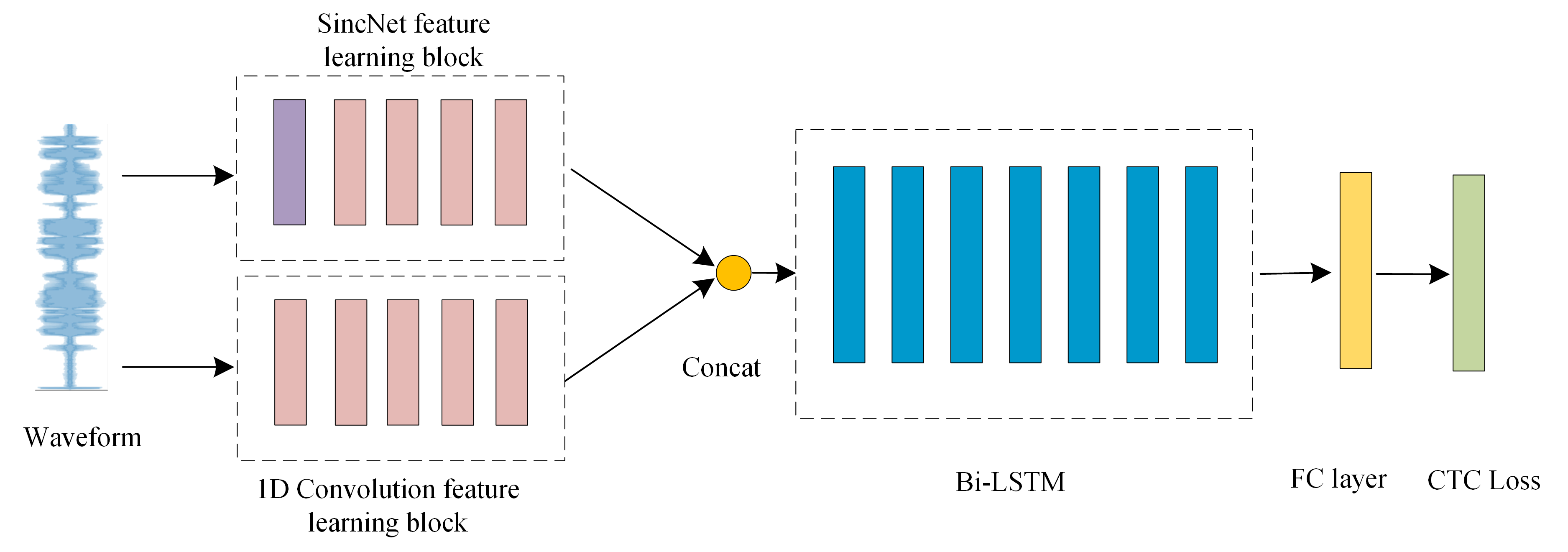}}
\caption{The architecture of the proposed end-to-end automatic speech  recognition system. The model consists of the feature learning block and the backbone network. A feature learning block is used to learn features from the raw waveforms.}
\end{figure*}

Recently, the end-to-end speech recognition system has provided higher performance than traditional methods for common ASR tasks \cite{kim2017joint}. However, current end-to-end ASR systems usually use mel-frequency cepstral coefficients (MFCCs) or filter-bank (FBANK) to process the raw waveform speech instead of directly inputting the raw waveform.

A deep learning-based feature extracted method--SincNet, was proposed to deal with ASR task and speaker recognition task, and the experimental results showed that the neural network based on SincNet achieved better performance for both the two tasks. The SincNet can learn more informative and discriminative features from raw waveform \cite{ravanelli2018speech}, \cite{ravanelli2018interpretable}. Other deep learning-based models, like wav2vec, were also proposed to extract speech feature from raw waveforms. The wav2vec model explores unsupervised pre-training for speech recognition by learning representations from raw audio through several 1D convolution layers. The wav2vec model is trained on large amounts of unlabeled speech and the resulting representations serve as the input of the acoustic model
for the ASR task \cite{schneider2019wav2vec}. 

In previous studies, the well-designed MFCC or FBANK features were applied to perform preliminary processing on the raw waveform. In this procedure, the raw speech is divided into frames with 25 ms frame length and 10 ms shift, and a series of signal processing transformations are applied to convert the 1D waveform into 2D feature map. After the raw waveform speech is processed by MFCC or FBANK, the extracted feature map is fed into the neural network for acoustic modeling. This method has achieved state-of-the-art results in many ASR tasks. However, the design of FBANK and MFCC is based on the human ear's response to audio, it may lose some of the raw waveform speech information. Considering the complex ATC environment, the handcrafted feature engineering may not be an optimal option for ASR tasks. Therefore, the learning mechanism was proposed to learn informative and discriminative features from raw waveforms, which achieved desired performance improvement for common ASR applications, such as SincNet, wav2vec \cite{ravanelli2018interpretable}, \cite{schneider2019wav2vec}.  

In this work, an end-to-end neural network is designed to achieve the ASR task in the ATC domain, in which a novel feature learning block is proposed to extract high-level speech representation from raw waveforms. Both the SincNet and 1D convolution block are designed to learn features from raw waveforms. The backbone network is constructed by cascading the convolutional neural network (CNN) and recurrent neural network (RNN) layers and is jointly optimized with features learning block by the connectionist temporal classification (CTC) loss function. Most importantly, the proposed model leverages the feature learning block to implement the end-to-end training, which predicts the text sequence from raw waveform without any pretraining. 

The model proposed in this paper introduces a feature learning approach for speech recognition tasks in the ATC domain, and the work in this paper is dedicated to solving the multilingual speech recognition problem in the ATC domain. The experimental results show that our approach outperforms other baselines on the ATCSpeech corpus, achieving a 6.9\% character error rate (CER), i.e., 0.9\% absolute CER reduction over other baselines.

\section{Related work}
\label{sec:Related work}

With the rapid development of deep learning techniques in the past decades, it has outperformed the conventional methods for the ASR task \cite{sainath2015convolutional,soltau2016neural}. 
For the common ASR applications, the SincNet was proposed to be combined with end-to-end architecture, which achieved higher accuracy on Wall Street Journal corpus \cite{parcollet2020e2e}. In addition, the sinc-convolution was combined with the depthwise convolutions to construct the lightweight sinc-convolutions (LSC) model \cite{kurzinger2020lightweight}, which serves as a learnable feature extraction block for end-to-end ASR systems with minor trainable parameters.
Furthermore, the wav2vec2.0 is combined with the BERT to construct an end-to-end ASR model, achieving higher performance by learned features from pre-trained models \cite{yi2021efficiently}.

For the ASR research in the ATC domain, several state-of-the-art ASR models were applied to build a benchmark, which was trained on more than 170 hours of ATC speech \cite{juan2020automatic}. The contextual knowledge was integrated into the ASR model to achieve semi-supervised training, which provided better performance for recognizing callsign of the ATC instruction \cite{zuluaga2021contextual}. Our previous work built a unified framework for multilingual speech recognition in the ATC system to translate ATC and pilot multilingual speech to text \cite{lin2020unified}. To deal with the scare of high quality annotated training data in the ATC domain, a novel method was proposed to leverage pretraining and transfer learning \cite{lin2021improving}.

\section{METHODLOGY}
\label{sec:method}

In this work, a complete end-to-end architecture is proposed to achieve the multilingual ASR task in the ATC domain by cascading a feature learning block and backbone network. Both the SincNet and convolutional layers are applied to formulate a feature learning block to extract high-level representations from raw waveforms. 
By combining with the backbone network (referred to the Deep Speech 2 model \cite{amodei2016deep}), the proposed model is finally optimized by the CTC loss function. In addition, to consider the multilingual ASR issue in the ATC domain, a special vocabulary is built based on the Chinese characters and English letters. The architecture of the proposed model is illustrated in Fig.1. Thanks to the design of the feature learning block, the proposed model is able to directly predict the text sequence from the waveforms and can be optimized in an end-to-end manner with the goal of an ASR task, instead of a pretraining task.

\subsection{Feature learning block}

In the ATC domain, the ASR task faces the challenges of multilingual language, complex background noise, etc, which results in the fact that the handcrafted feature engineering may not be an optimal option for the ASR task. Therefore, the feature learning block is applied to extract features from raw waveforms in a learnable way, which further supports the acoustic modeling of the ASR task.

In this work, a novel feature learning block is proposed to learn informative features from raw waveforms, whose outputs are generated by concatenating the feature maps of different learning paths. In general, the feature learning block consists of two paths:
a) SincNet path: including a sinc-covolutional layer and four convolutional layers. The configurations are as follows: filter-sizes [129, 3, 3, 3, 3], strides [1, 1, 1, 1, 1], max pooling size [3, 3, 3, 3, 3] and ReLU activations.
b) Common convolutional path: including four convolutional layers, whose configurations are the same as that of the SincNet path to keep the same feature map size, supporting the concatenation operation of the two paths.

SincNet is a novel convolutional neural network, it first layer is Sinc-convolution. The Sinc-convolution layer is an improved neural network based on the parametrized Sinc functions, which has the ability to learn task-oriented features from speech signal directly. In general, the convolution operation is defined as follows:
\begin{equation}
y[n]=x[n]*h[n]=\sum_{l=0}^{L-1}{x[l]\cdot h[n-l]}.
\end{equation}
\begin{equation}
y[n]=x[n]*g[n,\theta],
\end{equation}
where $*$ is convolution operation, $x[n]$ is a chunk of the speech signal, $h[n]$ is a filter with length $L$, and $y[n]$ is final filtered result. Compared to a large number of parameters in the common CNN block, the Sinc-convolution operator is able to achieve the signal processing with much fewer trainable parameters by defining the filter function $g$. In this work, the rectangular bandpass filter is applied to serve as the filter-bank to implement the speech processing, this function $g$ can be written in the time domain as follow:
\begin{equation}
g[n,f_1,f_2 ]=2f_2 sinc(2\pi f_2 n)-2f_1 sinc(2\pi f_1 n).
\end{equation}
In the above function, the $sinc$ function is defined as $sinc(x)=sin(x)/x$, where $f_1$ and $f_2$ are the learned low and high cutoff frequency of the bandpass filter,  respectively. The parameter $f$ is randomly initialized in the range of $[0,f_s/2]$, and $f_s$ is  the sample rate of the speech signal.
\begin{equation}
g_w [n,f_1,f_2 ]=g[n,f_1,f_2]\cdot w[n].
\end{equation}
\begin{equation}
w[n]=0.54-0.46cos(\pi n / L).
\end{equation}

Furthermore, to smooth out the abrupt discontinuities at the end of $g$, an available option is used to make $g$ multiplied by a window function $w$, such as the popular Hamming window \cite{ravanelli2018interpretable}.

\subsection{The backbone network}
Our work is motivated by the end-to-end ASR system Deep Speech 2  \cite{amodei2016deep}. In this work, we explore architecture with a feature learning block and backbone network. The backbone network is consists of 7 bidirectional long short-term memory (Bi-LSTM) layers and a fully connected layer,  which is further optimized with the CTC loss. The batch normalization is applied to speed up the model convergence, while the dropout layer is used to prevent the overfitting problem. The ReLU is selected as the activation function for the proposed model.

In the proposed end-to-end ASR model, the goal is to predict the text sequence $L = \{l_1,…,l_m\}$ from the input speech signal $X = \{x_1 ,…,x_n\}$, in which $l_i$ is from a special vocabulary based on Chinese characters and English letters.

 In general, multiple frames in $X$ correspond to a token of $L$. The length of speech frames is usually much longer than the label length. To address this issue, the CTC loss function was designed to achieve the alignment between the speech and label sequence automatically. The $t$ th frame corresponds to the output label $k$. Given the speech input $X$, the probability of the output sequence $\pi$ is shown in (6). Therefore, the probability of the final sequence can be obtained by (7), in which $w$ is the set of all possible sequences. For example, by using '$\_$' to denote a blank, both the outputs "$X\_YY\_Z$" and "${\_XY\_Z\_}$" correspond to the final output "$XYZ$" \cite{graves2006connectionist}.

\begin{equation}
p(\pi|X)=\prod_{t=1}^T{y_{\pi_t}^t}, \pi_t\in A.
\end{equation}
\begin{equation}
p(l|X)=\sum_{\pi\in w^{-1}(l)}p(\pi|X).
\end{equation}
\section{EXPERIMENTS}
\label{sec:experiments}
\subsection{ATC corpus}
In this work, the training data of the proposed model is the ATCSpeech corpus. The ATCSpeech corpus is manually annotated from a real ATC environment \cite{yang2019atcspeech}. The ATCSpeech corpus is a multilingual corpus containing Chinese and English speeches. 
There are 16939 transcribed English utterances (about 18.69-hour) in the corpus, while 45586 (about 39.83-hour) for Chinese speech. The division for train, validation, and test set can be found in Table 1.

\begin{table}[h]
\renewcommand\arraystretch{1}
\label{tab0}
\caption{Data size of the corpus. "\#U" denotes the speeches utterances and "\#H" denotes the speeches hours}
\setlength{\tabcolsep}{2.0mm}
\begin{tabular}{cccccccc}
\hline
\hline
\multirow{2}{*}{\textbf{Language}} & 
\multicolumn{2}{c}{Train} & \multicolumn{2}{c}{Dev} & \multicolumn{2}{c}{Test} \\ \cline{2-7} 
&\#U & \#H & \#U & \#H & \#U &\#H\% \\
\hline
Chinese & 43186 & 37.77 & 1200 & 1.04 & 1200 & 1.03 \\
English &	15282 &	16.84 & 850 & 0.95 & 807 & 0.89\\

\hline
\end{tabular}
\end{table}

\subsection{Experimental configurations}
In this work, the proposed model is constructed based on the open framework PyTorch 1.7.0. The training server was equipped with an Intel i7-9700 processor, a single NVIDIA TITAN RTX GPU, 32-GB memory, and an Ubuntu 18.04 operating system.

During the model training, the Adam optimizer is used to optimize the trainable parameters. The initial learning rate is 0.0001. The batch size is set to 32. In the first epoch, the speech samples are sorted in reverse order (based on speech duration) to detect the overflow of GPU memory as early as possible. In the following epochs, the training samples are shuffled to improve the model robustness. The vocabulary is built on the Chinese characters and English letters, and also with some special tokens ($< UNK >$, $< SPACE >$). Finally, a total of 705 tokens in the vocabulary. 
As in \cite{yang2019atcspeech}, the Deep Speech 2 \cite{amodei2016deep}, Jasper \cite{li2019jasper}, and Wav2Letter++ \cite{pratap2019wav2letter++} are selected as the baselines, which are applied to achieve the monolingual and multilingual ASR task in this work. In order to ensure the fairness of the experiment, all those models are trained on the same dataset (ATCSpeech) without extra training data. 

To confirm the effectiveness of the acoustic model, no language model is integrated into the ASR decoding procedure, i.e., greedy strategy.

\subsection{Overall results}

In this section, all the models are applied to achieve both monolingual and multilingual ASR tasks. To confirm the efficacy of the feature learning block, the handcrafted feature engineering is performed to generate the input for baseline models. The experimental results for all the models are reported in Table 2-4. 

In general, the proposed approach yields the highest performance among all the models for both monolingual and multilingual ASR tasks. In general, the Deep Speech 2 baseline obtains better performance among the three baselines. This can be attributed to the that the temporal modeling provides significant performance improvement for the ASR task, which is also the motivation of our backbone network. In addition, the multilingual ASR system is able to obtain higher accuracy than that of the monolingual ASR system, benefiting from the larger dataset and prominent discrimination between Chinese and English. 
Specifically, for the Chinese speech, the proposed model achieves 7.6\% CER, i.e., 0.5\% absolute CER reduction over the Deep Speech 2 model. For the English speech, the proposed model obtains 0.4\% absolute CER reduction over the Jasper model (highest baseline accuracy), and the final CER reaches 8.9\%. 
To be specific, for the multilingual ASR task, the proposed model achieves higher performance than that of the monolingual ASR system, achieving 6.9\% CER. The results also confirm the effectiveness of the feature learning block in the proposed model, which learns more informative and discriminative features for the ASR task.

\begin{table}[h]
\label{tab1}
\renewcommand\arraystretch{1}
\caption{The result of Chinese speech. “Fea.” details the type of input
features employed, and “Training” denotes the loss and epoch of the training. “Dev” denotes the validation dataset. Results are expressed in CER.}
\setlength{\tabcolsep}{0.9mm}
\begin{tabular}{ccccccc}
\hline
\hline
\multirow{2}{*}{\textbf{Models}} & \multirow{2}{*}{\textbf{Fea.}} & 
\multicolumn{2}{c}{Training} & Dev &Test \\ \cline{3-6}

                                          &       &
                                     Loss & Epoch & CER \%&CER\% \\
\hline
DS2 \cite{yang2019atcspeech} & FBANK & 0.53 & 33& 8.1&8.1 \\
Jasper10*3 \cite{yang2019atcspeech} &	FBANK &	2.45 & 101& 11.2 & 11.3\\
Wav2letter++ \cite{yang2019atcspeech} & FBANK & 2.37 & 136 & 14.2& 14.3\\
ours & RAW & 0.26 &	105 &7.6& 7.6\\
\hline
\end{tabular}
\end{table}

\begin{table}[h]
\renewcommand\arraystretch{1}
\label{tab2}
\caption{The result of English speech. }
\setlength{\tabcolsep}{0.9mm}
\begin{tabular}{ccccccc}
\hline
\hline
\multirow{2}{*}{\textbf{Models}} & \multirow{2}{*}{\textbf{Fea.}} & 
\multicolumn{2}{c}{Training} & Dev & Test \\ \cline{3-6}
                                          &       &
                                     Loss & Epoch &CER\% & CER\% \\
\hline
DS2 \cite{yang2019atcspeech} & FBANK & 0.54 & 107& 10.4& 10.4 \\
Jasper10*3 \cite{yang2019atcspeech} &	FBANK &	0.91 & 200 & 9.2 & 9.3\\
Wav2letter++ \cite{yang2019atcspeech} & FBANK & 1.06 & 307 & 11.3& 11.4\\
ours & RAW & 0.21 &	150 &8.9& 8.9\\

\hline
\end{tabular}
\end{table}

\begin{table}[!h]
\renewcommand\arraystretch{1}
\label{tab3}
\caption{The result of multilingual speech. }
\setlength{\tabcolsep}{1.4mm}
\begin{tabular}{ccccccc}
\hline
\hline
\multirow{2}{*}{\textbf{Models}} & \multirow{2}{*}{\textbf{Fea.}} & 
\multicolumn{2}{c}{Training} & Dev & Test \\ \cline{3-6}
                                          &       &
                                     Loss & Epoch& CER\% & CER\% \\
\hline
DS2 & FBANK & 0.45 & 30 & 7.8 & 7.8 \\
Jasper10*3 &	FBANK &	2.35 & 97 &10.0& 10.1\\
Wav2letter++ & FBANK & 2.29 & 115 &12.3& 12.4\\
ours & RAW & 0.26 &	102 &6.9& 6.9\\

\hline
\end{tabular}
\end{table}

\subsection{Ablation study}

\begin{table}[h]
\renewcommand\arraystretch{1}
\label{tab4}
\caption{The result of the model with different SincNet kernel sizes. "Kernel Size" details the kernel size of the proposed method first convolution layer}
\setlength{\tabcolsep}{2.0mm}
\begin{tabular}{cccccc}
\hline
\hline
\multirow{2}{*}{\textbf{Kernel Size}} & 
\multicolumn{2}{c}{Training} &Dev & Test \\ \cline{2-5}
                                          &      
                                     Loss & Epoch & CER\%& CER\% \\
\hline
 251 & 0.31 & 90 &7.3& 7.4 \\
 129 &	0.26 & 102 &6.9& 6.9\\
 65 & 0.25 & 95 &7.0& 7.1\\

\hline
\end{tabular}
\end{table}

In this section, we explore different convolution configurations of the SincNet by ablations to find an optimal model architecture. As illustrated in \cite{ravanelli2018interpretable}, the size of the convolution kernel and the size of max pooling will affect the risk of aliasing in the filtered signal. The kernel size of 251 (as in \cite{ravanelli2018interpretable}) is selected as the baseline for the proposed model in this work.

Considering the higher speech rate in the ATC environment \cite{lin2020unified}, a smaller kernel size is designed for the proposed feature learning block (for both the SincNet and common CNN paths), including 129 and 65. For the convolutional operation, a smaller kernel takes fewer frames as a single phoneme state, corresponding to a higher speech rate. The experimental results are listed in Table 5. It can be seen that the proposed model achieves the highest accuracy with the kernel size of 129, which is the optimal option for the ATCSpeech corpus.

In addition, the contribution of the different paths in the feature learning block is also considered to improve the final performance.
As listed in Table 6, a total of 4 configurations are designed in this ablative study, including the single path (a and b) and two paths (c and d). As can be seen from the experimental results, both the four configurations achieve desired performance improvement baseline models, i.e., 7.4\% CER v.s. 7.8 \% CER, which confirms the effectiveness of the learning mechanism for the ASR task in the ATC domain.
In general, the two paths configuration achieves higher performance than that of the single path, which benefits from the model capacity for learning diverse task-oriented features. In addition, the single path SincNet (a) is able to obtain comparable performance with that of the two paths SincNet (c), which indicates that only simply accumulating the same architecture fails to learn informative features for improving the ASR performance. 
Finally, the proposed model achieves the most significant performance improvement (6.9\% CER) by concatenating the SincNet and CNN block, which support the motivation of model design in this work. In conclusion, both the model capacity and the feature diversity are indispensable to tackle the ASR specificities in the ATC domain.

\begin{table}[!h]
\renewcommand\arraystretch{1}
\label{tab5}
\caption{The results of different feature learning networks Structure. "Models" is the backbone network with different feature learning block}
\setlength{\tabcolsep}{2.0mm}
\begin{tabular}{ccccc}
\hline
\hline

\multirow{2}{*}{\textbf{Models}} & \multicolumn{2}{c}{Training}&Dev & Test \\ \cline{2-5}
                                          &     
                                     Loss & Epoch & CER\% & CER\% \\
\hline
1D CNN*1 (a)  & 0.27 &	93 &7.4& 7.5\\
SincNet*1 (b)  & 0.26 & 95 &7.1 & 7.2 \\
SincNet*2 (c)  &	0.25 &97 & 7.1& 7.1 \\
ours (d)  & 0.26 &	102 &6.9 & 6.9 \\

\hline
\end{tabular}
\end{table}

\section{CONCLUSION}
\label{sec:conclusion}

In this work, we propose using the feature learning method for end-to-end ASR tasks in the ATC domain, which results in better recognition performances without extra training data. The proposed method allows direct end-to-end training. For the ASR in the ATC domain, the proposed feature learning block can learn more meaningful information from raw waveforms, and the feature learning-based ASR system performance beyond the handcrafted feature learning-based ASR system. Moreover, the two-path feature learning block can learn more diverse features than that of the single-path feature learning ASR system in the field of ATC. In addition, it is also found that adding SincNet paths cannot further improve the final performance, while better results can be obtained by cascading SincNet and 1D CNN feature learning blocks. Thanks to the increased capacity of the proposed model, more diverse features can be learn in ASR tasks in the ATC domain. The experimental results have demonstrated the proposed method outperforms other baselines in multilingual ASR tasks on ATCSpeech corpus.


\vfill

\newpage
\newpage
\bibliography{reference_google}

\begin{thebibliography}{10}

\bibitem{el2011survey}
Moataz El~Ayadi, Mohamed~S Kamel, and Fakhri Karray,
\newblock ``Survey on speech emotion recognition: Features, classification
  schemes, and databases,''
\newblock {\em Pattern Recognition}, vol. 44, no. 3, pp. 572--587, 2011.

\bibitem{geacuar2010reducing}
Claudiu-Mihai Geac{\u{a}}r,
\newblock ``Reducing pilot/atc communication errors using voice recognition,''
\newblock in {\em Proceedings of ICAS}, 2010, vol. 2010.

\bibitem{lin2020unified}
Yi~Lin, Dongyue Guo, Jianwei Zhang, Zhengmao Chen, and Bo~Yang,
\newblock ``A unified framework for multilingual speech recognition in air
  traffic control systems,''
\newblock {\em IEEE Transactions on Neural Networks and Learning Systems},
  2020.

\bibitem{lin2019real}
Yi~Lin, Linjie Deng, Zhengmao Chen, Xiping Wu, Jianwei Zhang, and Bo~Yang,
\newblock ``A real-time atc safety monitoring framework using a deep learning
  approach,''
\newblock {\em IEEE Transactions on Intelligent Transportation Systems}, vol.
  21, no. 11, pp. 4572--4581, 2019.

\bibitem{kim2017joint}
Suyoun Kim, Takaaki Hori, and Shinji Watanabe,
\newblock ``Joint ctc-attention based end-to-end speech recognition using
  multi-task learning,''
\newblock in {\em 2017 IEEE International Conference on Acoustics, Speech and
  Signal Processing (ICASSP)}. IEEE, 2017, pp. 4835--4839.

\bibitem{ravanelli2018speech}
Mirco Ravanelli and Yoshua Bengio,
\newblock ``Speech and speaker recognition from raw waveform with sincnet,''
\newblock {\em arXiv preprint arXiv:1812.05920}, 2018.

\bibitem{ravanelli2018interpretable}
Mirco Ravanelli and Yoshua Bengio,
\newblock ``Interpretable convolutional filters with sincnet,''
\newblock {\em arXiv preprint arXiv:1811.09725}, 2018.

\bibitem{schneider2019wav2vec}
Steffen Schneider, Alexei Baevski, Ronan Collobert, and Michael Auli,
\newblock ``wav2vec: Unsupervised pre-training for speech recognition,''
\newblock {\em arXiv preprint arXiv:1904.05862}, 2019.

\bibitem{sainath2015convolutional}
Tara~N Sainath, Oriol Vinyals, Andrew Senior, and Ha{\c{s}}im Sak,
\newblock ``Convolutional, long short-term memory, fully connected deep neural
  networks,''
\newblock in {\em 2015 IEEE International Conference on Acoustics, Speech and
  Signal Processing (ICASSP)}. IEEE, 2015, pp. 4580--4584.

\bibitem{soltau2016neural}
Hagen Soltau, Hank Liao, and Hasim Sak,
\newblock ``Neural speech recognizer: Acoustic-to-word lstm model for large
  vocabulary speech recognition,''
\newblock {\em arXiv preprint arXiv:1610.09975}, 2016.

\bibitem{parcollet2020e2e}
Titouan Parcollet, Mohamed Morchid, and Georges Linares,
\newblock ``E2e-sincnet: Toward fully end-to-end speech recognition,''
\newblock in {\em ICASSP 2020-2020 IEEE International Conference on Acoustics,
  Speech and Signal Processing (ICASSP)}. IEEE, 2020, pp. 7714--7718.

\bibitem{kurzinger2020lightweight}
Ludwig K{\"u}rzinger, Nicolas Lindae, Palle Klewitz, and Gerhard Rigoll,
\newblock ``Lightweight end-to-end speech recognition from raw audio data using
  sinc-convolutions,''
\newblock {\em arXiv preprint arXiv:2010.07597}, 2020.

\bibitem{yi2021efficiently}
Cheng Yi, Shiyu Zhou, and Bo~Xu,
\newblock ``Efficiently fusing pretrained acoustic and linguistic encoders for
  low-resource speech recognition,''
\newblock {\em IEEE Signal Processing Letters}, vol. 28, pp. 788--792, 2021.

\bibitem{juan2020automatic}
Zuluaga-Gomez Juan, Petr Motlicek, Qingran Zhan, Rudolf Braun, and Karel
  Vesely,
\newblock ``Automatic speech recognition benchmark for air-traffic
  communications,''
\newblock Tech. {R}ep., ISCA, 2020.

\bibitem{zuluaga2021contextual}
Juan Zuluaga-Gomez, Iuliia Nigmatulina, Amrutha Prasad, Petr Motlicek, Karel
  Vesel{\`y}, Martin Kocour, and Igor Sz{\"o}ke,
\newblock ``Contextual semi-supervised learning: An approach to leverage
  air-surveillance and untranscribed atc data in asr systems,''
\newblock {\em arXiv preprint arXiv:2104.03643}, 2021.

\bibitem{lin2021improving}
Yi~Lin, Qin Li, Bo~Yang, Zhen Yan, Huachun Tan, and Zhengmao Chen,
\newblock ``Improving speech recognition models with small samples for air
  traffic control systems,''
\newblock {\em Neurocomputing}, vol. 445, pp. 287--297, 2021.

\bibitem{amodei2016deep}
Dario Amodei, Sundaram Ananthanarayanan, Rishita Anubhai, Jingliang Bai, Eric
  Battenberg, Carl Case, Jared Casper, Bryan Catanzaro, Qiang Cheng, Guoliang
  Chen, et~al.,
\newblock ``Deep speech 2: End-to-end speech recognition in english and
  mandarin,''
\newblock in {\em International conference on machine learning}. PMLR, 2016,
  pp. 173--182.

\bibitem{graves2006connectionist}
Alex Graves, Santiago Fern{\'a}ndez, Faustino Gomez, and J{\"u}rgen
  Schmidhuber,
\newblock ``Connectionist temporal classification: labelling unsegmented
  sequence data with recurrent neural networks,''
\newblock in {\em Proceedings of the 23rd International Conference on Machine
  Learning}, 2006, pp. 369--376.

\bibitem{yang2019atcspeech}
Bo~Yang, Xianlong Tan, Zhengmao Chen, Bing Wang, Dan Li, Zhongping Yang, Xiping
  Wu, and Yi~Lin,
\newblock ``Atcspeech: a multilingual pilot-controller speech corpus from real
  air traffic control environment,''
\newblock {\em arXiv preprint arXiv:1911.11365}, 2019.

\bibitem{li2019jasper}
Jason Li, Vitaly Lavrukhin, Boris Ginsburg, Ryan Leary, Oleksii Kuchaiev,
  Jonathan~M Cohen, Huyen Nguyen, and Ravi~Teja Gadde,
\newblock ``Jasper: An end-to-end convolutional neural acoustic model,''
\newblock {\em arXiv preprint arXiv:1904.03288}, 2019.

\bibitem{pratap2019wav2letter++}
Vineel Pratap, Awni Hannun, Qiantong Xu, Jeff Cai, Jacob Kahn, Gabriel
  Synnaeve, Vitaliy Liptchinsky, and Ronan Collobert,
\newblock ``Wav2letter++: A fast open-source speech recognition system,''
\newblock in {\em ICASSP 2019-2019 IEEE International Conference on Acoustics,
  Speech and Signal Processing (ICASSP)}. IEEE, 2019, pp. 6460--6464.

\end{thebibliography}

\bibliographystyle{IEEEbib}

\end{document}